\documentclass[review]{elsarticle}
\usepackage{setspace}
\usepackage{graphicx}
\usepackage{amsmath}
\usepackage{float}
\usepackage{hyperref}
\usepackage{cleveref}

\newcommand{\1}{$\langle100\rangle$}
\newcommand{\2}{$\langle110\rangle$}
\newcommand{\3}{$\langle111\rangle$}

\title{Comparison of SIA Defect Morphologies from Different Interatomic Potentials for Collision Cascades in W}

\author[addbarcv]{Utkarsh Bhardwaj}
\ead{butkarsh@barc.gov.in}

\author[addhelsinki,addaalto]{Andrea E. Sand}
\ead{andrea.sand@helsinki.fi}

\author[addbarcv,addhbni]{Manoj Warrier}
\ead{manojwar@barc.gov.in}

\address[addbarcv]{Computational Analysis Division, BARC, Vizag, AP,
India-530 012}
\address[addhelsinki]{Department of Physics, P.O. Box 43, FI-00014 University
of Helsinki, Finland}
\address[addaalto]{Department of Applied Physics, Aalto University, FI-00076 Aalto, Espoo, Finland}
\address[addhbni]{Homi Bhabha National Institute,
Anushaktinagar, Mumbai, Maharashtra, India - 400 094}

\begin{document}

\begin{abstract}
The morphology of defects formed in collision cascades is an essential aspect of the subsequent evolution of the microstructure. The morphological composition of a defect decides its stability, interaction, and migration properties.  We compare the defect morphologies in the primary radiation damage caused by high energy collision cascades simulated using three different interatomic potentials in W. An automated method to identify morphologies of defects is used. While most defects form 1/2\3 dislocation loops, other specific morphologies include \1 dislocation loops, multiple loops clustered together, rings corresponding to C15 configuration and its constituent structures, and a combination of rings and dislocations. The analysis quantifies the distribution of defects among different morphologies and the size distribution of each morphology. We show that the disagreement between predictions of the different potentials regarding defect morphology is much stronger than the differences in predicted defect numbers.
\end{abstract}

\maketitle

\section{Introduction} \label{sec:intro}

Predictive simulations of irradiation effects on material properties can be divided into two broad multi-scale studies: (i) Modeling the change in the microstructure of materials due to irradiation and (ii) Modeling the change in material properties due to the change in the microstructure. The first step in modeling the microstructure change is quantifying the number of defects, their spatial and size distributions, and classifying the defect cluster morphologies due to collision cascades initiated by energetic primary knock-on atoms (PKA). This information can then be used as input to higher scale simulations that model the migration, interaction and morphological evolution of these defects \cite{heinisch1992molecular, becquart2000primary, SOUIDI2001179, bukonte2013comparison, de2016primary, BACON20001, SINGH1997107, OSETSKY200065, BECQUART200639, OSETSKY2002852, GAO2000213}. The morphology of a defect and the arrangement of its constituent components influence its thermal stability, migration properties, and interaction with other defects. Therefore it is necessary to classify the different possible defect morphologies occuring during primary damage in order to simulate micro-structure evolution.

Molecular dynamics (MD) simulations are widely used to simulate the primary damage caused by collision cascades \cite{PhysRev.120.1229, PhysRev.133.A595, PhysRev.139.A118, Stoller2012293, nordlund1995molecular, cai20121}. MD simulations give a good insight into the atomistic mechanisms of defect formation and clustering. They have been used to propose new formulae for the number of defects created based on physically realistic damage models \cite{KaiNatureComm2018}. In addition to the defect concentration, size distribution of point defect clusters and percentage of point defects in clusters have also been widely studied \cite{Stoller2012293, warrier2015statistical}. The initial studies of defect morphologies were limited to the qualitative assessment of defects produced \cite{OSETSKY200185, BACON20001} and their properties \cite{OSETSKY200065, GAO2000213}. With the increase in simulation data and automatic data-driven techniques, it is now possible to identify the morphology of all the defects produced and analyze them quantitatively \cite{BHARDWAJ2020109364, bhardwaj2020graph}. MD simulations can use different interatomic potentials developed for the same material. The assessment of variability in simulation results arising from the choice of different potentials is an essential step to determine the uncertainity of results before using them in higher scale models \cite{OSETSKY200185, BYGGMASTAR2018530}.

The number of surviving point defects produced in W collision cascades shows notable variation across different interatomic potentials \cite{SAND2016119, SETYAWAN2015329}.  The comparative study of W inter-atomic potentials by Sand et al. shows that for energies up to 50 keV, the number of point defects correlates well with the threshold displacement energy and the form of potential in the short-range repulsive part \cite{SAND2016119}. However, for higher energy cascades, the number of surviving point defects has been shown to only correlate well with the size of clusters formed for different interatomic potentials. Another earlier study by Sand et. al. \cite{Sand_2013} uses individual inspection guided by potential energy analysis to report qualitative differences in dislocation loops produced with different interatomic potentials in W at 150 keV. The study shows that for the potential by Derlet et al. \cite{PhysRevB.76.054107}, half of the smaller loops have Burgers vector 1/2\3  and other half have \1, whereas half of all the larger clusters formed complex sessile configurations with sometimes visible partial loops. With the Ackland-Thetford EAM potential \cite{A-T}, all the clusters having more than size 30 formed loops with 1/2\3 Burgers vector with the exception of one \1 loop. Experiments have verified the presence of both \3 and \1 loops \cite{yi2012}. A later study by Wahyu et al. \cite{SETYAWAN2015329} using the Ackland-Thetford potential, finds similar morphologies with clusters bigger than size 30 mostly forming \3 loops while all the clusters with size less than 30 are classified as 3D clusters that do not form loops. The study also shows a few \1 loops, mixed loops, and 3D clusters of sizes greater than 30. The exact morphology of 3D clusters is not clear. The qualitative differences in defect morphologies formed with the different potentials and the presence of complex configurations or ambiguous 3D clusters demand a closer quantitative assessment of the defect morphologies formed with various potentials in W. Quantitative analysis of defect morphologies, other than giving essential insights into the radiation damage at atomistic scales, also helps in (a) providing inputs to higher scale models, (b) understanding the sessile/glissile nature of different defects, (c) understanding the interaction between different defect types (d) understanding their stability and morphological transitions, and (e) development of interatomic potentials that result in the defect morphologies that are consistent with experiments and Density Functional Theory (DFT) based studies \cite{PhysRevB.94.024103, ALEXANDER2020152141}. 

We present a comparative study of interatomic potentials of W based on the defect morphology formed in the high energy collision cascades. The method is applied to study and compare point defect clusters formed in W collision cascades at 100 keV and 200 keV using three different potentials. The potentials include the Finnis-Sinclair potential \cite{doi:10.1080/01418618408244210} as modified by Juslin et al. (JW) \cite{JUSLIN201361}, potential by Derlet et al. \cite{PhysRevB.76.054107} with the repulsive part fitted by Bj\"orkas et al. \cite{BJORKAS20093204} (DND-BN), and the potential by Marinica et al. \cite{Marinica_2013}, stiffened for cascade simulations by Sand et al. (M-S) \cite{SAND2016119}. The three potentials have been earlier compared for other parameters such as defect count, and cascade evolution behavior \cite{SAND2016119}. The aim of this study is to classify the various defect morphologies that are formed in collission cascades in W resulting from three widely used interatomic potentials and discuss the differences in results from these potentials with regard to cluster morphologies.

We classify the defect morphologies into dislocation loops with their Burgers vector orientation, C15 like 3D rings and their basis structures \cite{DEZERALD2014219, PhysRevLett.108.025501, bhardwaj2020graph} that also form independent defects \cite{}, mixed dislocation loops with details of their constituent loops, and defects composed of both ring and dislocation loops. The morphological analysis is done using a graph theory \cite{tarjan1972connected} based method that resolves every point defect cluster into its homogeneous constituent components and characterizes them \cite{bhardwaj2020graph}.  We compare the distribution of defects in different cluster morphologies and the size distribution of each morphology separately. We show that the disagreement between predictions of the different potentials regarding defect morphology is more substantial than the differences in predicted defect numbers.

\section{Methods}
\label{sec:Methods}

The MD simulations of collision cascades for 100 and 200 keV PKA were performed using the PARCAS MD code \cite{parcas}. A primary knock-on atom (PKA) was selected from among the lattice atoms of a cubic simulation cell and given the desired kinetic energy in a random initial direction. Periodic boundaries were used for each cascade. Simulations in which atoms with kinetic energy above 10 eV crossed any of the periodic boundaries were aborted, and the initial position of the recoil was shifted further from the border. Temperature control at 0K was applied to all atoms within three atomic layers from the cell borders using a Berendsen thermostat \cite{doi:10.1063/1.448118}. Each cascade was followed until the cell had cooled to an average temperature of only a few K, ensuring stable defects.  Electronic stopping was considered, with a lower velocity cutoff of 10 eV, using the stopping power tables from SRIM \cite{srim}.

The defect morphology identification is carried out using Savi, a recent graph-based morhology identification method \cite{bhardwaj2020graph, bhardwajcsaransh}. The method resolves the homogenous constituent components of a defect and characterizes it based on the morphology of the components. This enables a clear definition of mixed morphology defects. A component comprises dumbbells/crowdions that hold a specific relationship with their neighboring dumbbells/crowdions. The method identifies various properties of dumbbells/crowdions, such as their extents, orientations, distances, and angles with neighbors to identify and characterize the components. For example, the angle of neighboring dumbbells/crowdions in a dislocation loop is close to zero, while in a C15 like ring, it is sixty degrees. The method identifies the Burgers vector direction and magnitude of each component loop in a defect. It also can classify a C15 like a ring and its constituent basis shapes even in the presence of a trapped dislocation loop. The method can be extended to identify any specific defect geometry and different materials.

\section{Results}
\label{sec:Results}

\subsection{Defect Morphologies in W collision cascades}

\begin{figure}[ht]
  \centerline{\includegraphics[width=.95\linewidth]{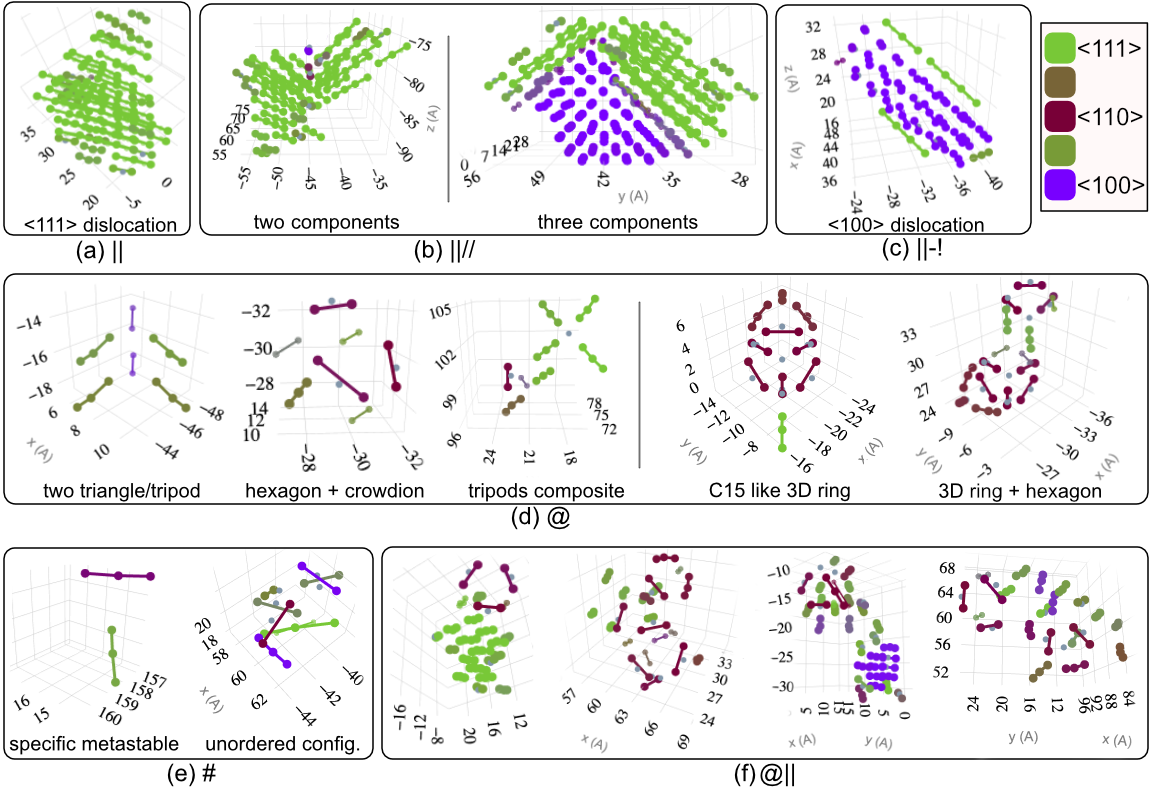}}
  \caption{\label{fig:fig-1}
    Defect morphology formed in the W collision cascades along with the symbols used to represent them. (a) parallel bundle of \3 directed SIA that form 1/2\3 edge dislocation loop, (b) defects composed of multiple dislocation loops (here three loops, 2 in \3 and 1 in \1 direction), (c) parallel group of \1 directed SIA that form \1 loop, (d) 3D C15 like rings and their constituent basis, (e) non-parallel and non-ring configurations (f) defects with dislocations trapped with rings (here hexagonal planar ring along with \3 loop). Lines are drawn along the dumbbells/crowdions and are colored according to their orientation as specified in the legend (top-right).
  }
\end{figure}

\Cref{fig:fig-1} shows the different morphologies found in the W collision cascades. A bundle of parallel dumbbells/crowdions form a dislocation loop with Burgers vector direction similar to the dumbbells/crowdion orientation. The first three morphologies in (a), (b), and (c) are formed of dislocation loops. The \1 loop almost always exhibits a few \3 crowdions on the periphery, as shown in (c). Multiple dislocation loops compose together to form a sessile defect. Another stable sessile defect morphology is C15-like ring morphology, as shown in (d). The constituent basis structures that a C15 ring is composed of, viz. triangle/tripod and hexagonal ring can exist as a separate stable sessile defect \cite{DEZERALD2014219, PhysRevLett.108.025501, BHARDWAJ2020109364, bhardwaj2020graph}. A pair of orthogonal dumbbells/crowdions is another specific structure that is observed relatively rarely (the first defect in \Cref{fig:fig-1} (e)). A small fraction of defects also settle down transiently in a configuration with no specific order of dumbbells/crowdions (the second defect in \Cref{fig:fig-1} (e)). These are primarily non-recurring transient variations of parallel or sometimes ring morphologies appearing different due to thermal vibrations. A defect can be composed of both a dislocation loop and a ring, as shown in \Cref{fig:fig-1} (f).  
\begin{table}[H]
    \centering
    \caption{Description and ASCII symbol for different cluster morphologies.}
    \label{table:table-1} 
    \begin{tabular*}{0.99\textwidth}{@{\extracolsep{\fill}} | c | l |}
      \hline
      Symbol & Description\\
      \hline
      $||$  & Parallel bundle (dislocations) in \3 orientation \\ \hline
      $||-!$  & Parallel bundle (dislocations) in \1 orientation (having \3 \\
              & crowdions on fringes)\\ \hline
      $||//$  & Multi-component dislocations \\ \hline
      $@$  & Rings, (C15-like or its basis shapes) \\ \hline
      $@||$  & Composed of both, rings \& dislocations\\ \hline
      \#  & Non-parallel \& non-ring configurations \\ \hline
    \end{tabular*}
\end{table}

\Cref{table:table-1} lists the different morphologies and the symbol used to represent them. The symbols are chosen such that they indicate the morphology.

\subsection{Distribution of defects in different cluster morphologies}
\label{subsec1:Results}

The kind of defect morphologies found for all the three interatomic potentials are broadly the same. The morphologies include dislocation loops in \3 and \1 orientations, rings corresponding to C15-like configuration and its constituent basis, and a few non-parallel and non-ring transient configurations. However, the proportions and sizes of the morphologies differ substantially.

\begin{figure}[ht]
  \centerline{\includegraphics[width=.9\linewidth]{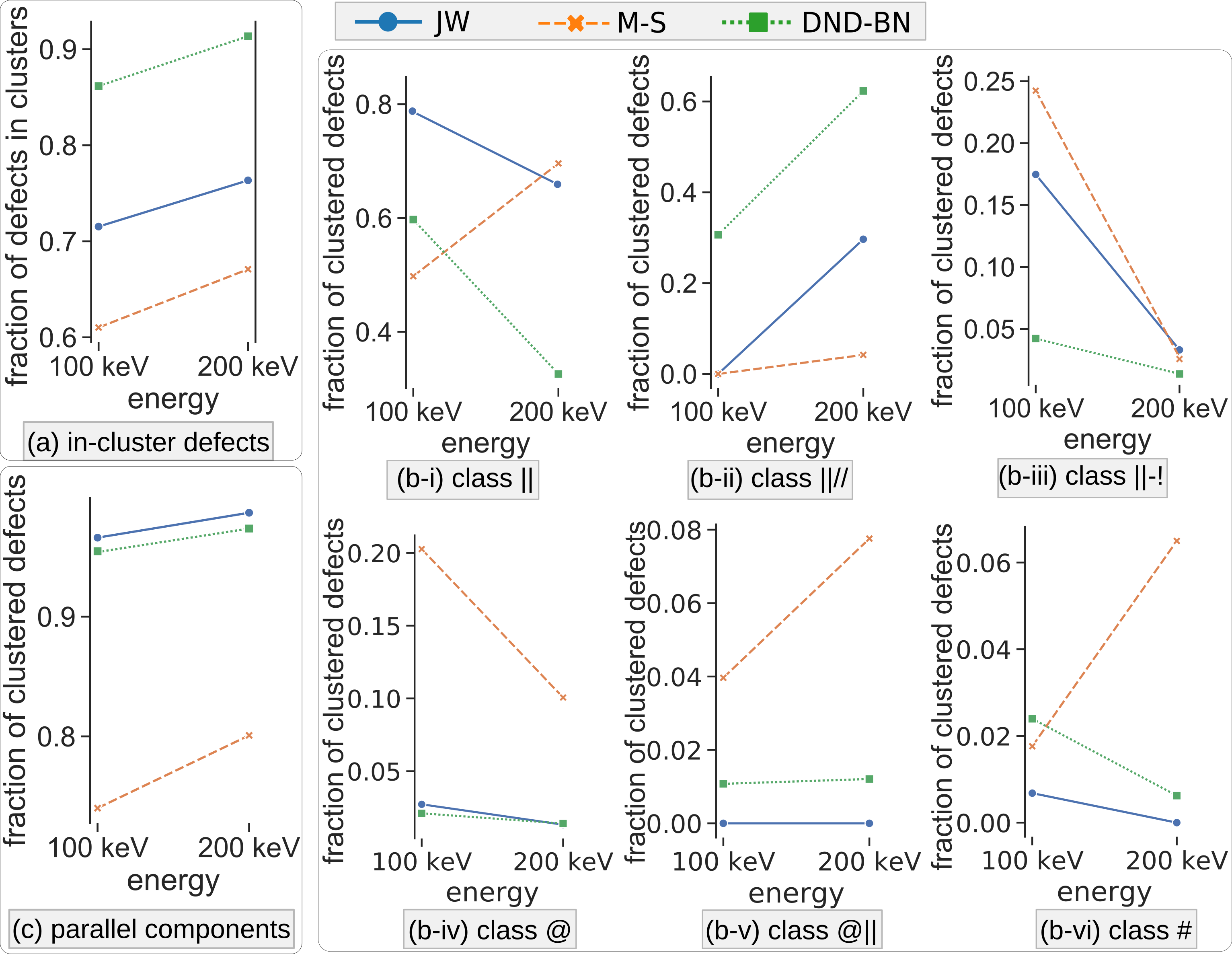}}
  \caption{\label{fig:fig-2} (a) Shows the fraction of in-cluster defects for 
  the three potentials (represented by different style and colors; blue: JW, 
  orange: M-S, green: DND-BN). The fraction increases with energy for all the three potentials (b) shows the fraction of in-cluster defects that are in parallel components that form dislocation loops. These include single dislocations in \3 and \1 directions as well as multi-component dislocations. (c) shows the distribution of the in-cluster 
  defects in different morphologies viz. (i) glissile single parallel SIAs 
  component in \2 orientation ($||$), (ii) multiple parallel
  components ($||//$) (iii) single parallel component in \3
  orientation ($||-!$), (iv) rings ($@$), (v) parallel and ring combination ($@||$), (vi) non-parallel, non-ring configurations (\#). The cluster shapes for different  morphologies are shown in \Cref{fig:fig-1}.
}
\end{figure}

The M-S potential shows more preference for forming rings as opposed to DND-BN and JW that prefer parallel components forming dislocation loops (\Cref{fig:fig-2} b). DND-BN shows a greater preference for multi-component dislocation loops ($||//$) than JW potential which prefers \3 dislocation loops ($||$) (\Cref{fig:fig-2} c-i, c-ii).

\begin{figure}[H]
  \centerline{\includegraphics[width=.5\linewidth]{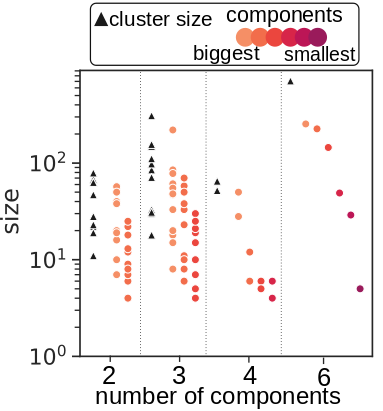}}
  \caption{\label{fig:fig-3}
The number of components and their sizes for for multi-component dislocation loops ($||//$). The x-axis shows the count of the components such that all the clusters with the same number of components are plotted together. The components are arranged in descending order of their sizes and are represented with a color gradient.}
\end{figure}

The fraction of defects forming $||$ morphology drops with energy in JW and DND-BN but rises in M-S (\Cref{fig:fig-2} c-i). The overall fraction in parallel components still increases for all the three potentials (\Cref{fig:fig-2} b) as the PKA energy increases. The drop in  the $||$ and $||-!$  morphology in DND-BN and JW is compensated by the stark increase in $||//$ morphology (\Cref{fig:fig-2} b-ii). The distribution of the number of components in $||//$ morphology and their sizes are shown in \Cref{fig:fig-3}.

\begin{figure}[ht]
  \centerline{\includegraphics[width=.5\linewidth]{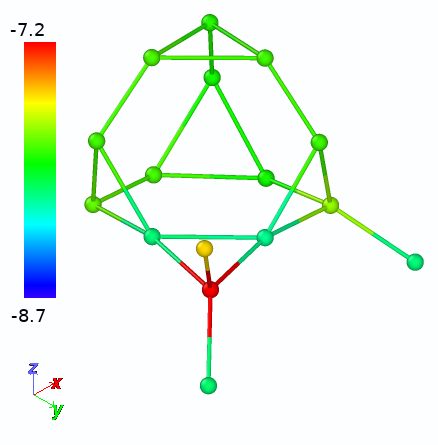}}
  \caption{\label{fig:fig-4} A cluster found at 100keV showing a C15 like
  structure consisting of four hexagons joined at 12 vertices forming a Laves Polyhedron with a crowdion tail appended to it. The color map shows the potential energy of the atoms.}
\end{figure}

The structure of 3D rings corresponds to the C15 symmetry is shown in \Cref{fig:fig-4}. Out of all the rings for each potential, less than $10\%$ are C-15 like 3D rings, and the rest are composed of planar ring bases (hexagonal ring and triangle/tripod).
 
\Cref{table:table-2} and \Cref{table:table-3} lists the distribution of the absolute number of point defects and defect clusters, respectively. Note that the number of cascades is different for different potentials and energies in the data-set, and the absolute numbers must be compared, keeping the number of cascades in context. The relative numbers per cascade have been shown in \Cref{fig:fig-2} earlier. The number of clusters formed for each potential is well above 100. The total number of clusters is 574.

\begin{table}[ht]
    \centering
    \caption{Distribution of absolute number of point defects. $E_{pka}$ is in keV.}
    \label{table:table-2} 
    \begin{tabular*}{0.99\textwidth}{@{\extracolsep{\fill}} | l l | l | l | l | l | l | l | l | l |}
      \hline
      Potential& $E_{pka}$  & cascades & $||$ & $||-!$ & $||//$ & $@$ & $@||$ & \# & total\\ \hline
      JW     & 100        & 5        & 242  & 42     &  0     & 8   & 0     & 2  & 294  \\
             & 200        & 5        & 459  & 11     &  202   & 9   & 0     & 0  & 681  \\
      M-S    & 100        & 5        & 113  & 55     &  0     & 46  & 9     & 4  & 227  \\
             & 200        & 5        & 296  & 45     &  20    & 48  & 37    & 31 & 477  \\
      DND-BN & 100        & 20       & 1226 & 109    &  622   & 43  & 22    & 23 & 2045 \\
             & 200        & 9        & 845  & 50     &  1584  & 36  & 31    & 16 & 2562 \\
      \hline
    \end{tabular*}
\end{table}

\begin{table}[H]
    \centering
    \caption{Distribution of absolute number of defect clusters. $E_{pka}$ is in keV.}
    \label{table:table-3} 
    \begin{tabular*}{0.99\textwidth}{@{\extracolsep{\fill}} | l l | l | l | l | l | l | l | l | l |}
      \hline
       & $E_{pka}$ & cascades & $||$ & $||-!$ & $||//$ & $@$ & $@||$ & \# & total\\ \hline
      JW     & 100        & 5        & 34   & 2      &  0     & 4   & 0     & 1  & 41   \\
             & 200        & 5        & 76   & 1      &  2     & 4   & 0     & 0  & 83   \\
      M-S    & 100        & 5        & 40   & 2      &  0     & 14  & 2     & 1  & 59   \\
             & 200        & 5        & 61   & 2      &  1     & 20  & 4     & 5  & 93   \\
      DND-BN & 100        & 20       & 127  & 5      &  15    & 11  & 3     & 9  & 170  \\
             & 200        & 9        & 92   & 2      &  11    & 14  & 3     & 6  & 128  \\
      \hline
    \end{tabular*}
\end{table}

\subsection{Cluster size distribution across cluster morphologies}
\label{subsec2:Results}

\begin{figure}[H]
  \centerline{\includegraphics[width=.9\linewidth]{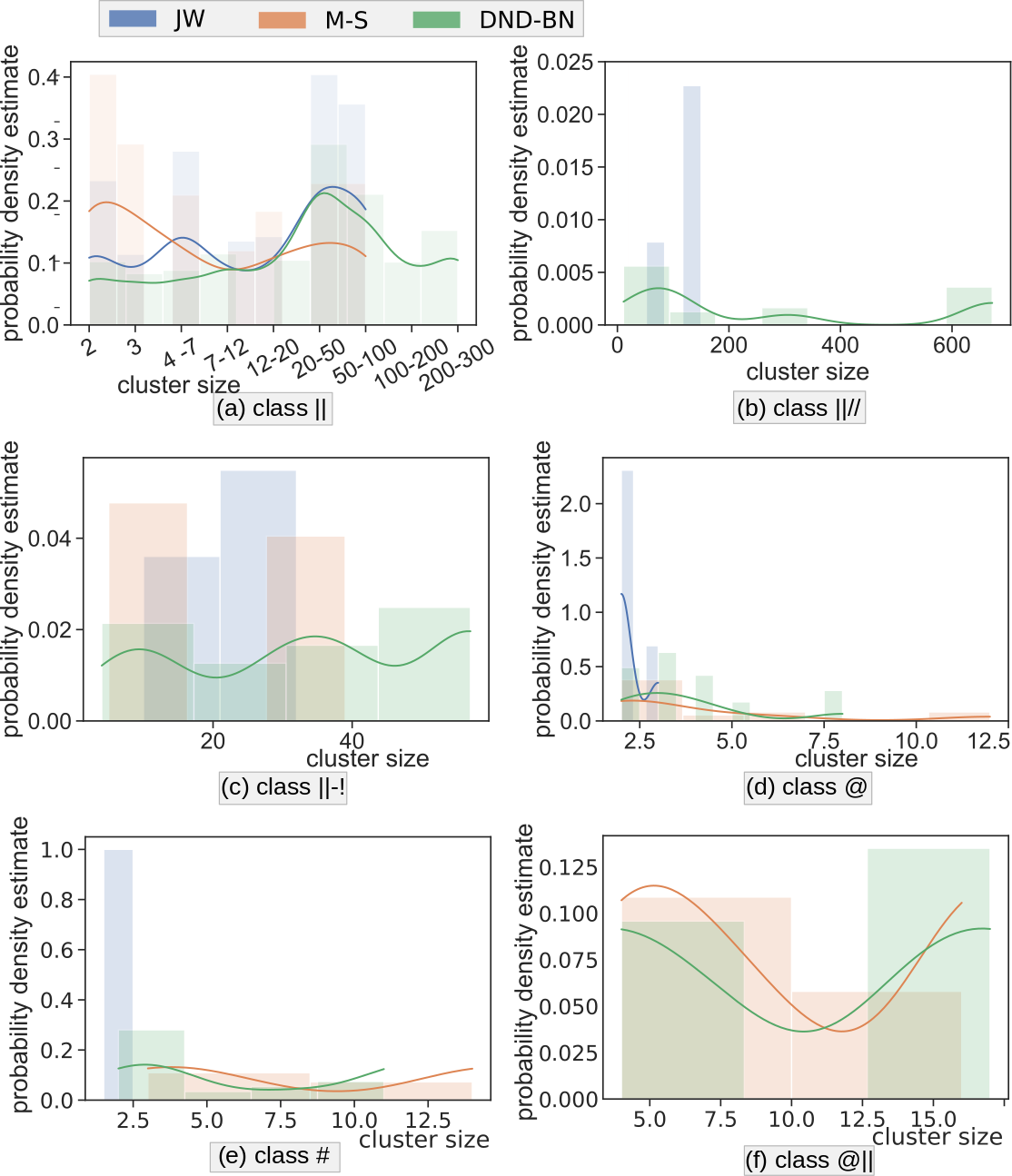}}
  \caption{\label{fig:fig-5}
  Cluster size distribution for a different class of morphologies. The y-axis
  shows the probability density for different defect sizes marked on the x-axis. The bars and curves show histogram bin values and probability estimates, respectively. The curves are only drawn for cases with sufficient data. Different potentials are shown with different colors.
  }
\end{figure}

\Cref{fig:fig-5} shows the size distribution of defects for different
morphologies. The values are probability density estimates for each potential. In general, the defects with non-parallel morphology are small when compared to the parallel dislocations ((a), (b), and (c) \Cref{fig:fig-5}). The DND-BN potential has large clusters with parallel components in classes $||$, $||-!$ and especially $||//$. The biggest cluster is formed with six parallel components, out of which three are oriented in \2 and three in \3.

For the M-S potential, most of the clusters in the $||$ class are of size two. For non-parallel classes, the size distribution is spread out. For JW potential, there are no unordered arrangements, the only cluster that appears in class \#, is a pair of orthogonal dumbbells(the first defect in \Cref{fig:fig-1} (e)). For class $@$ in JW, mostly all the defects are of size two while a few also have size three.

\subsection{Deviation from primary alignments and lattice site non-collinearity}
\label{subsec5:Results}

For a single SIA dumbbell or crowdion, or for parallel bundles of dumbbells and crowdions, it is expected that the lattice site is collinear with the line that passes through the two SIA atoms that co-occupy it. The SIA dumbbell/crowdion is normally characterized as aligned to one of the primary orientations \2, \1
or \3. Recent DFT studies show that an SIA adopts a symmetry-broken
configuration in chromium, molybdenum, and tungsten
\cite{PhysRevMaterials.3.043606}. For non-parallel dumbbell configurations like basis of rings such as tripod, we observe that the crowdions are aligned slightly off from these standard orientations. Moreover, the deviations from the standard orientations are expected to be present due to thermal vibrations.

\begin{figure}[ht]
  \centerline{\includegraphics[width=.9\linewidth]{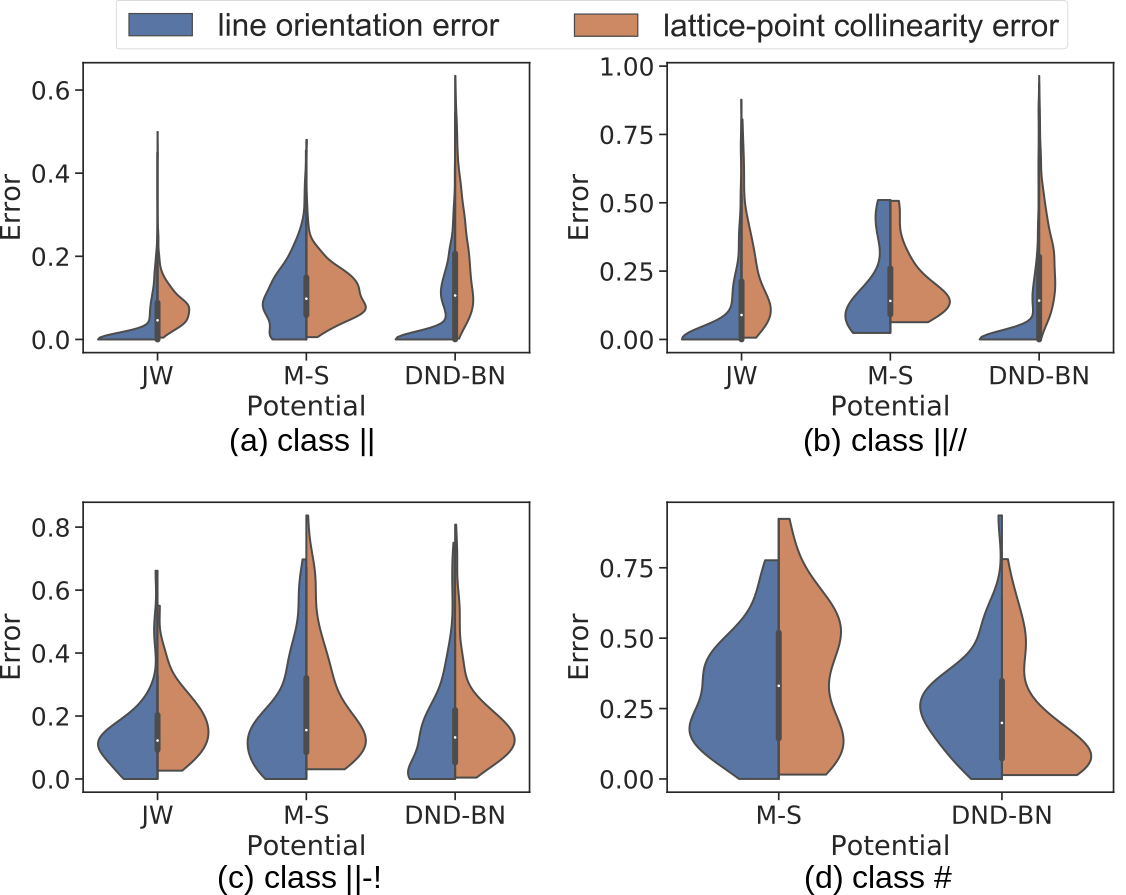}}
  \caption{\label{fig:fig-6}
  Statistical distribution of lattice point non-collinearity (red) and
  deviation from primary orientation (blue) for different classes across
  potentials. The deviations are lower for classes where parallel SIAs are
  more. For non-parallel random orientation class (\#) plotted in (d), the
  deviations are large as expected. Among potentials JW shows the lowest
  deviations especially for \2 dislocation loops ($||$).
  }
\end{figure}

\Cref{fig:fig-6} shows these two deviations viz. lattice-point
non-collinearity and deviation from perfect crowdion orientation for the
different classes and potentials. The lattice-point non-collinearity value is
taken as the perpendicular distance (in $\AA$) of the lattice point from the line defined by the two interstitials occupying it. The angular deviation from perfect crowdion orientation is normalized by dividing the angle with the maximum value of deviation observed for the data, i.e., 30 degrees.

The deviation in orientation is least in parallel \2 clusters ($||$), followed
by $||-!$ and $||//$. For non-parallel random orientation class (\#) plotted in (d), the deviation is large. The non-collinearity of lattice point follows the same trend, except $||//$ has slightly higher values than $||-!$. The deviation in orientation seems to be depending on the orientation of the SIA, with the least deviation in morphologies with \2 orientations, followed by \3 and \1. The non-collinearity can be viewed as increasing with the increase in SIAs having non-parallel neighborhood. For $||//$, non-parallel SIAs are present where two components are interfaced, however in $||-!$ it occurs when the \3 SIAs that are in main component are interfaced with surrounding non-parallel SIAs on the fringes.

Among potentials, JW shows the lowest deviations, especially for \2 dislocation loops ($||$). This behavior by JW is consistent with a strong preference for clusters with parallel SIAs and almost negligible preference for another arrangement of SIAs except for small di-interstitials, be it $@$ or \#.

\section{Discussion}

The energy available for defect formation during the collision cascade far
exceeds the formation energy of any quasi-stable defects in the observed
cluster-size range. Defect formation within the heat spike region \cite{San17}
thus becomes a question of the available time and transition pathways from the
disordered liquid-like state to the various lower energy configurations of
defects or perfect crystal. All interatomic potentials in this study are found
to produce a fraction of defect clusters which differ from the predicted lowest
energy defect configurations. The validity of this general result is supported by
experimental evidence indicating that \1 dislocation loops form directly
as a result of cascades in W, despite being energetically unfavorable with
respect to \3 loops \cite{Yi15}.

The relaxation process of the cascade core depends on the recrystallization
rate as well as on details of the potential energy landscape. From this
perspective, the M-S potential, which has been fitted using liquid
configurations in addition to perfect crystal and point defects, may capture
the relaxation process more accurately. Thus, the prediction of small C15-like
structures in the primary damage according to M-S may be significant, and
deserves further study, despite the fact that the C15 phase is energetically
unfavorable in tungsten \cite{Ale16}. On the other hand, the M-S potential
predicts a melting point far in excess of the experimental melting point of
tungsten \cite{SAND2016119}, and this has a strong impact on the
recrystallization rate \cite{Nor97f}, which may in turn lead to incorrect
predictions. In addition, the pronounced lack of large dislocation loops in the
damage predicted by the M-S potential appears to contradict experiment, where
impacts of 150 keV W ions are found to directly produce defects, up
to sizes of a few nanometers in diameter \cite{Yi15} which can be identified using a Transmission Electron Microscopy (TEM).

It is possible that real materials behave according to a blend of the
predictions of these different potentials. Since cascade-induced defects do
not necessarily form in lowest energy configurations, the predictions of the
cascade damage formation process cannot be validated through comparison with
low-energy defect structures computed e.g. with DFT. Resolving the
question of the accuracy of the various potentials for cascade simulations
most likely hinges on obtaining parameters suggested by each potential to carry out higher scale simulations, coupled with developments in
the resolution of experimental imaging techniques, since most of the clusters formed in-cascade are below the observational limit of conventional transmission electron microscopy \cite{San17}. 
\section{Conclusions}
\label{sec:conclude}

We have used a novel computational method based on graph theory to characterize the defect cluster morphologies in the primary radiation damage in tungsten as predicted by molecular dynamics simulations. The newly developed analysis method provides a detailed characterization of the morphology-based on the constituent homogeneous components of the defect clusters. We have applied the method to classify the morphologies of defects formed at energies of 100 keV and 200 keV, comparing the predictions of three different interatomic potentials. 

We find a marked difference in the predicted morphology of the primary damage simulated with the different potentials. In particular, we note that the JW potential displays the strongest tendency to form ideal dislocation loops with single Burgers vectors of either \3 or \1 direction, whereas the DND-BN potential additionally predicts the frequent formation of very large multi-component loops, where different parts have different Burgers vectors. The M-S potential, on the other hand, predicts a large fraction of defects that do not display a dislocation loop character, but rather contain components of 2D or 3D rings, including small clusters with a C15 structure. These clusters are typically much smaller than the well-formed dislocation loops, and would be below the resolution limit of TEM.

In line with the observation of the tendency to form well resolved defects, we also find that the JW potential predicts constituent SIA configurations most closely aligned with the bcc lattice positions. Nevertheless, all potentials show the formation of defect clusters with morphologies different from that which can be expected based solely on a minimum formation energy criteria for the respective potential.

The detailed investigation of the morphological structure presented here
provides a rigorous comparison of the predictions of different
interatomic potentials going beyond the conventional measure of the point defect count. Applied to larger databases, such as that currently being developed by the IAEA containing cascade damage configurations in various materials \cite{CascadesDB, cnnmeeting6}, this method can serve to inform larger-scale models designed to treat radiation damage evolution on longer time scales, where the morphology of defects has a significant impact on the stability, interaction and mobility of defects. As such, this work demonstrates a valuable tool in the ongoing effort to formulate multi-scale models of radiation effects in materials with increasing accuracy and predictive power.

\section*{References}
\bibliographystyle{elsarticle-num} 
\bibliography{PotentialComparison}

\end{document}